# AN INTELLIGENT MOBILE-AGENT BASED SCALABLE NETWORK MANAGEMENT ARCHITECTURE FOR LARGE-SCALE ENTERPRISE SYSTEM

A.K. Sharma[1], Atul Mishra[2], Vijay Singh[3]

[1]Department of Computer Engineering, YMCA University of Science and Technology, Haryana, India
`ashokkale2@rediffmail.com`
[2]Department of Computer Engineering, YMCA University of Science and Technology, Haryana, India
`mish.atul@gmail.com`
[3]Department of Computer Engineering, YMCA University of Science and Technology, Haryana, India
`vijaypunnu222@gmail.com`

## ABSTRACT

*Several Mobile Agent based distributed network management models have been proposed in recent times to address the scalability and flexibility problems of centralized (SNMP or CMIP management models) models. Though the use of Mobile Agents to distribute and delegate management tasks comes handy in dealing with the previously stated issues, many of the agent-based management frameworks like initial flat bed models and static mid-level managers employing mobile agents models cannot efficiently meet the demands of current networks which are growing in size and complexity. Moreover, varied technologies, such as SONET, ATM, Ethernet, DWDM etc., present at different layers of the Access, Metro and Core (long haul) sections of the network, have contributed to the complexity in terms of their own framing and protocol structures. Thus, controlling and managing the traffic in these networks is a challenging task. This paper presents an intelligent scalable hierarchical agent based model for the management of large-scale complex networks to address aforesaid issues. The cost estimation, carried out with a view to compute the overall management cost in terms of management data overhead, is being presented. The results obtained thereafter establish the usefulness of the presented architecture as compare to centralized and flat bed agent based models.*

## 1. INTRODUCTION

The new technologies and concepts in both data and telecommunication industry are dramatically changing the way enterprises provide, maintain and use the various IT services. Management systems have to accomplish the profound effects of this evolution that will introduce into the network enormous quantities of different network elements ranging from low resource devices to large scale distributed applications. The network management [1] scene is further complicated by distributed, often mobile, data, resources, service access and control, especially when these networks are growing in size and complexity [2][3]. Moreover, varied technologies, such as SONET, ATM, Ethernet, DWDM etc., present at different layers of the Access, Metro and Core (long haul) sections of the network, have contributed to the complexity in terms of their own framing and protocol structures. Thus, controlling and managing the traffic in these networks is a challenging task.

Existing management models traditionally adopt a centralized, Client/Server (C/S) approach wherein the management application, with the help of a manager, acting as clients, periodically accesses the data collected by a set of software modules, agents, placed on network devices by using an appropriate protocol as shown in figure 1.

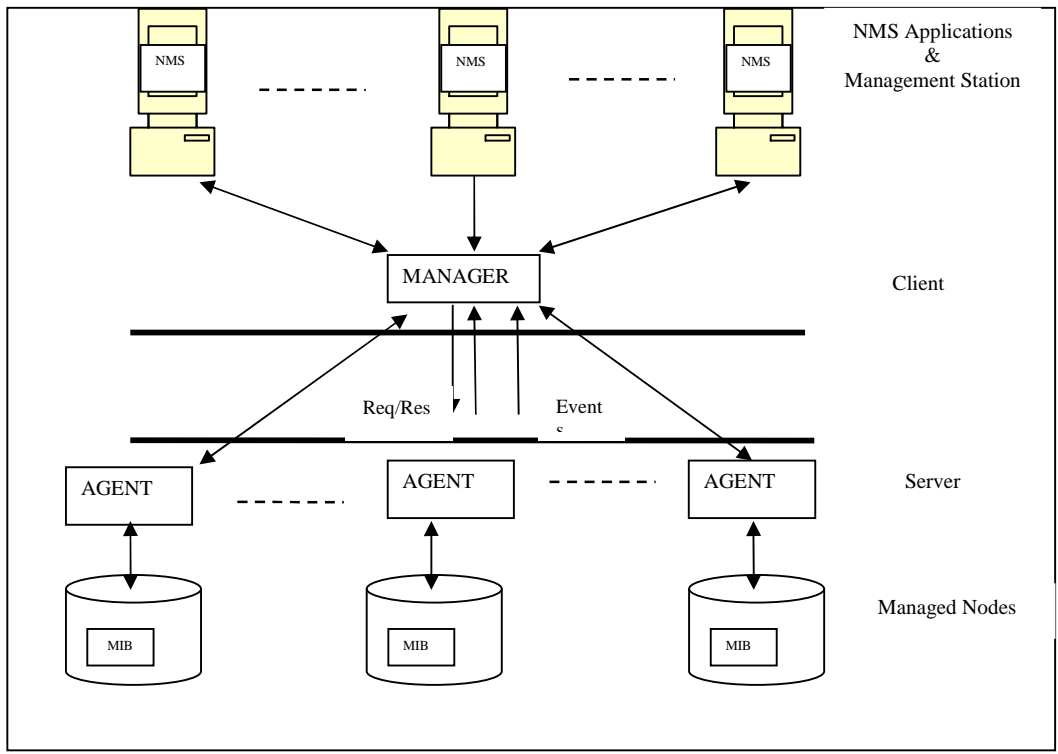

**Figure 1: Centralized Manager/Agent interaction**

The functionality of both managers and the agents where managers act as client and the agents as a server is statically defined at design time as shown in figure 2(a). Traditionally, the Network Management (NM) scene has been dominated by the Internet Engineering Task Force (IETF) simple network management protocol SNMP for data networks [4] and the OSI common management information protocol (CMIP) for telecommunication network [5], which are typically designed according to a centralized model and hence suffer from lack of distribution, a low degree of flexibility, low scalability and fault tolerance [6][7]. They also require network operators at NMS level to make real-time decisions and manually find solutions for the series of problems in the network. These network management systems deal only with data gathering and reporting methods, which in general involve substantial transmission of management data thereby consuming a lot of bandwidth and computational overhead. Moreover it also causes a considerable strain on the network and significant traffic jam at the manager host [8]. Besides this centralized management activities are limited in their capability as they cannot do intelligent processing like upfront judgment, forecasting, analyzing data and make positive efforts to maintain quality of service.

These problems have motivated a trend towards distributed management intelligence that represents a rational approach to overcome the limitations of centralized NM. As a result, several distributed management frameworks have been proposed both by researchers and standardization bodies [13][14]. However, these models are typically identified by static management components that cannot adapt to the evolving nature of today's networks, with rapidly changing traffic patterns and topology structures.

Of-late, the Mobile Agent (MA) paradigm has emerged within the distributed computing field. The term MA refers to autonomous programs with the ability to move from host to host to resume or restart their execution and act on behalf of users towards the completion of a given task [15]. One of the most popular topics in MA research community has been distributed NM, wherein MAs have been proposed as a means to balance the burden associated with the processing of management data and decrease the traffic associated with their transfers (data can be filtered at the source). Network management based-on Mobile agent [9][10][11][12] refers to equipping agents with network management intelligence and allowing them to issue requests to managed devices/objects after migrating close to them as depicted in figure 2(b).

The independence and mobility of mobile agents reduce client server bandwidth problems by moving a query from client to the server. It not only saves repetitive request/response handshake but also addresses the much needed problems created by intermittent or unreliable network connections. Agents can easily work off-line and communicate their results when the application is back on-line. Moreover, agents support parallel execution (load balancing) of large computation which can be easily divided among various computational resources.

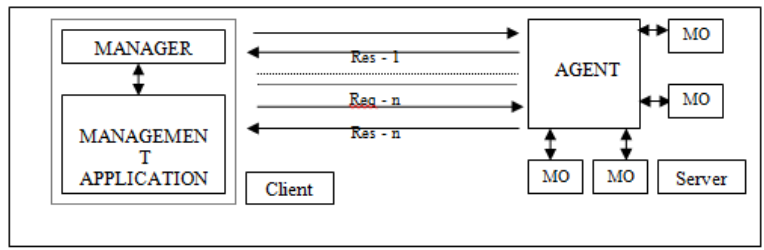

Figure 2(a): Client/Server design for information retrieval

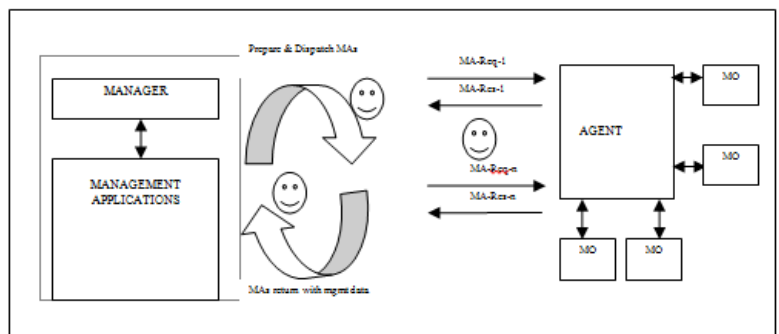

Figure 2(b): Mobile Agents for information retrieval

Keeping in view the rational presented above, we propose a sub-network layer partitioned network management model, called IMASNM, based-on mobile agents in order to minimize management data flow to a centralized server.

## 2. HIGH LEVEL DESIGN OF IMASNM MODEL

In this section we propose an architecture which provides a framework for Network Management functionality and related mobile agent management. Typical network management model based-on mobile agent works like this: management administrator dispatches corresponding mobile agents to managed nodes to perform different tasks, and these agents take back management information. Some shortcomings exist in this model like difficult to manage large-scale complex network, devices which do not support mobile agent run time environment escape from being managed and mostly it is incompatible with conventional SNMP based network management system.

Considering above, we proposed a Sub-network layer partitioned network management model shown in figure 3. In this scheme, the managed network is divided into many sub-networks based on the geographical layout of the network or number of nodes a manager can efficiently manage or average load on the entire network or certain kind of administrative relationship. In each sub-network, a Mobile-Subnetwork Layer Manager (M-SNLM) is appointed and for the overall managed network a set of various management applications along with a network manager act as a Global Network Manager (GNM). M-SNLMs, depending upon the growth in their domain, spawn additional child M-SNLMs for scalable management of the network. Newly spawned M-SNLMs take over a portion of the network from their parent M-SNLMs and act as local managers for that portion of the network for all management needs as shown in figure 4. For a large & complex network, in between M-SNLMs (appointed by GNM) and Network elements, there gets formed a hierarchy of supervisors (child M-SNLMs appointed by first level of M-SNLMs) which interact with immediate next level subordinate M-SNLMs and also manage a set of managed devices, initially assigned to them. Leaf level M-SNLMs are responsible for managing local network only. The management results are reported to Global NM via supervisors and Global NM is responsible for administrative sub-network partition, assigning tasks to each M-SNLM, indirectly, via supervisors and managing the whole network. Supervisors would maintain aggregated information such as available ports, wavelengths, states, fault reports and performance data about certain number of M-SNLMs and network management applications at NM layer would interact with these high level supervisory-managers instead of establishing direct connection with every network element. Provisioning operations (like Service, Equipment, LightPath) would be issues to high-level managers which then based on SubNetwork partitioning would delegate the information to subordinate managers. Monitoring of alarms and performance reports would operate in the reverse direction: M-SNLMs would aggregate it for their portion of the network and then would hand it over to their supervisors.

A management application would need to communicate with high-level supervisors in order to manipulate and monitor the entire network. In order to support the dynamic redistribution of the managers, so that they could adapt to the evolving nature of today's networks, with rapidly changing traffic patterns and topology structures, all the managers are mobile agents in themselves. Communication between managers takes place via deglets [16]. A deglet is a lightweight agent with a transient task based life cycle model. There are two types of deglets: Provisioning deglets are the ones which flow from Supervisors (including NM) to Subordinates and upward flowing, EventReporting deglets, are those which report fault and performance events. Thus we have a hierarchically distributed deployment of cooperating mobile agents or "managers" which would lead to significantly reduced processing requirement at the Network Management level.

Based on this, we propose a sub-network layer partitioned network management model, called IMASNM, based-on mobile agent in order to minimize management data flow to a centralized server. Intelligent agent allocating on managed devices performs local network management and reports the results to the superior manager, then the manager performs global network management using those submitted management results.

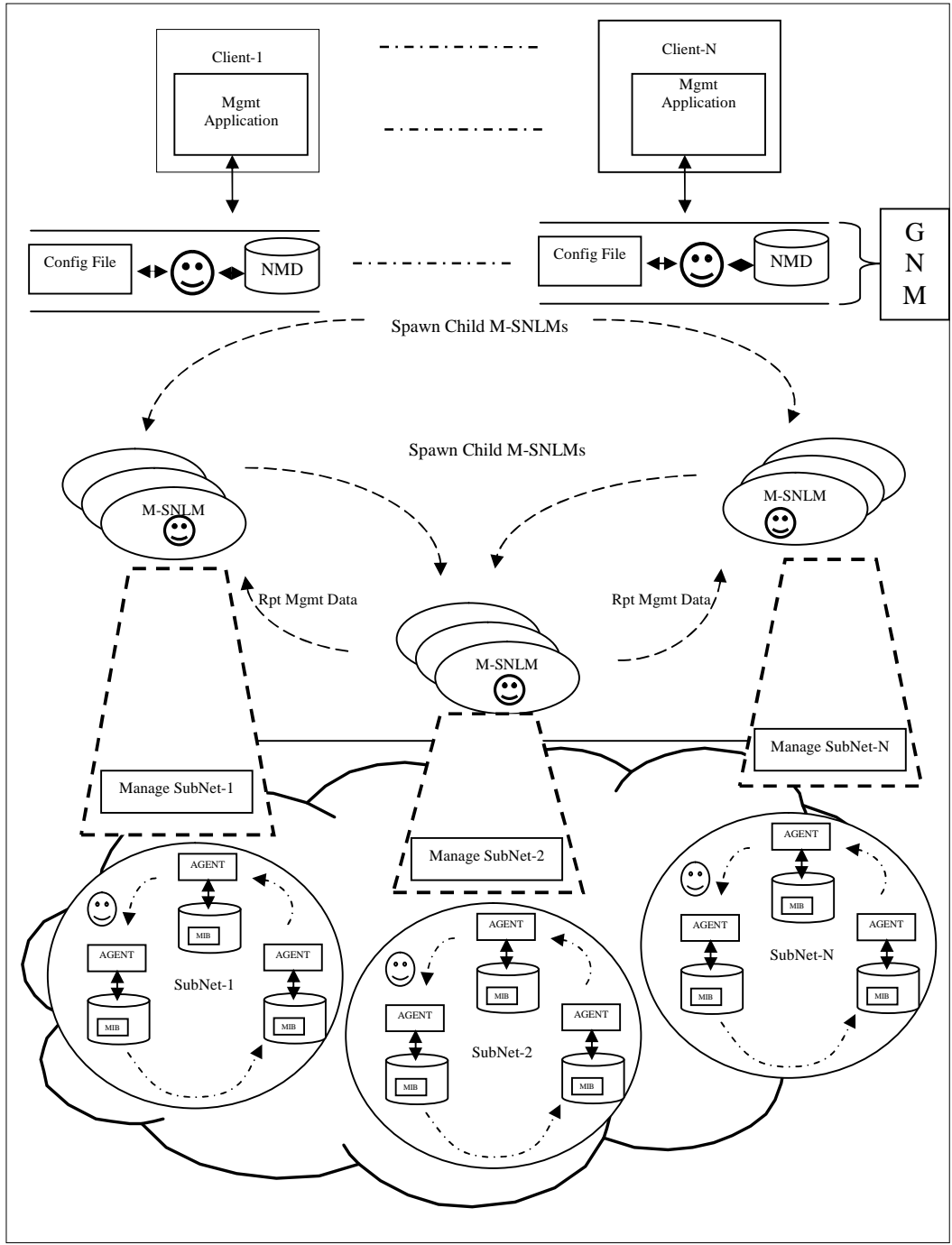

Figure 3: High Level Design of IMASNM

In proposed hierarchal network management model the mobile manager (M-SNLMs) move in their domains and manages network device with a flat bed model [11] scheme. The manager works with quite good efficiency if the size of domain falls in a specified range. If the domain size increases from that specified range then efficiency of network manager goes down which is the main problem of flat bed model.

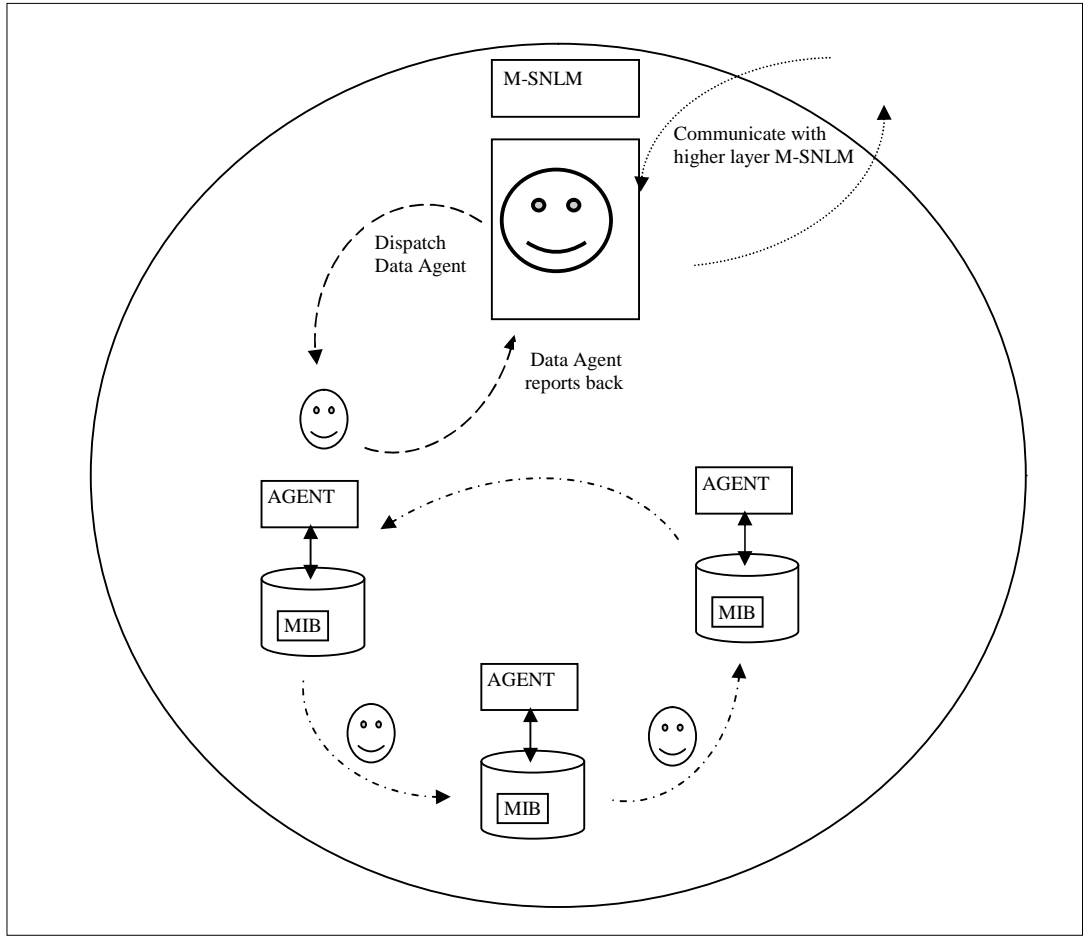

Figure 4. M-SNLM Managing a SubNet

This problem is solved in IMASNM model by using a simple concept, i.e., if the size of the managed domain becomes greater than a specified maximum range, then mobile manager divides manage domain and creates its own child and assign responsibility of newly created domain to its child. The child is a clone of original mobile manager having same management parameters and a maximum domain size equal to the mother mobile manager which the child manager can manage efficiently. As the child manager is a clone of mother manger it is also capable of creating its own children if its managing domain size increases. This phenomenon continues until whole network is managed with mobile managers with efficiently.

## 3. WORKING OF IMASNM MODEL

To understand the working of the IMASNM, consider an initial network of size N (say10 nodes) which is shown in figure 5. Let us consider the maximum size of a domain, which a mobile manager can efficiently manage, to be $M_{max}$ (say 3 nodes). After the initial discovery of network as network management operation based on IMASNM model starts, the central manager divides the network of N (here 10) nodes into four sub domains, out of which, three sub domains are of maximum manageable domain size i.e. $M_{max}$ nodes and one domain of size one node. For instance as show in figure 6 there are three sub domains consists of 3 nodes (10/3) each and one sub domain of size 1. It then creates three children (sub-domain manager) and assigns each one to manage one of the three sub-domains.

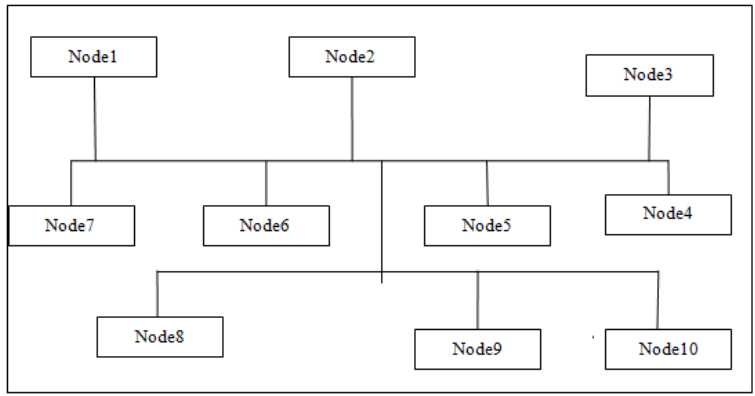

Figure 5. Initial Unmanaged Network View

It may be noted that the 4[th] domain i.e D1 (of size one node) is managed by the central manager itself whereas the other domains are being managed by separate managers. Then whole network is managed by using different level of intelligent manager hierarchy as shown in figure 7.

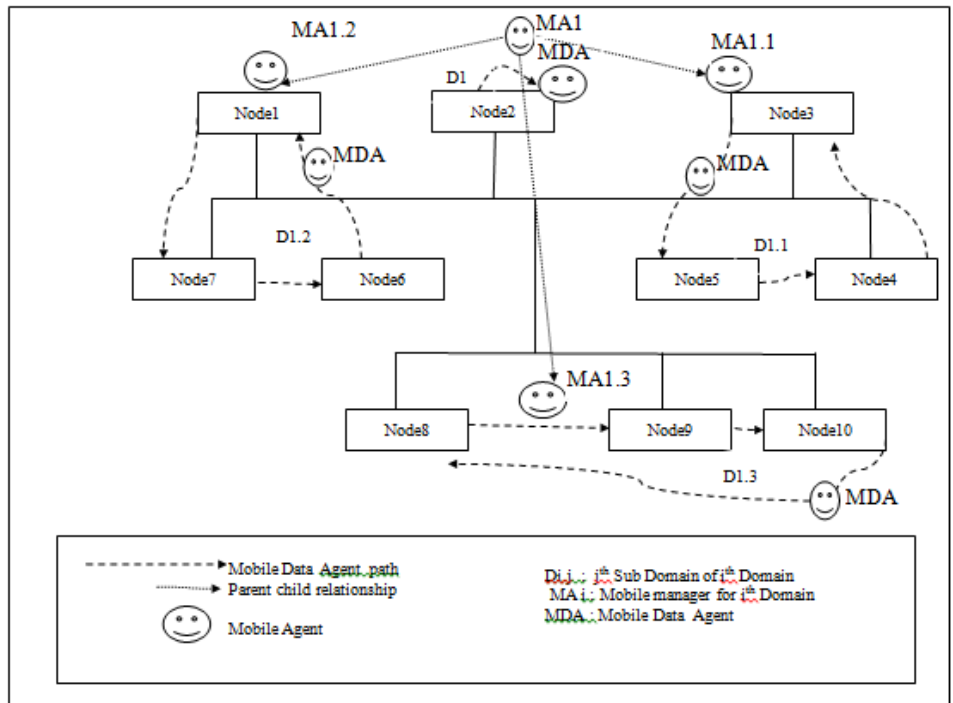

Figure 6. The Managed Network View based on ISHMANM Model

The manager $MA_x$ i.e manager of domain $D_x$ manages its domain $D_x$ as well as takes care of any increase in the size of $D_x$ due to addition of new network nodes. If there are additions in a domain (sub domain) and its size becomes greater then $N_{max}$ then the domain (sub domain) manager divides its domain into further sub domains of size less then equals to $N_{max}$ and generate its own clone to managed these sub domains. As shown in figure 7 the domain size of D1.3 domain becomes five due to discovery of nodes11and node12. As a result manager MA1.3 divides D1.3 into D1.3 and D1.3.1 domains and generated its clone i.e MA1.3.1 for managing domain D1.3.1. On the similar grounds domains D1.2 becomes D1.2 and D1.2.1 also D1.3.1 becomes D1.3.1 and D1.3.1.1 due to addition of Node 14-16 and Node 17-19 respectively into domains D1.2 and D1.3.1 which is shown in figure 8.

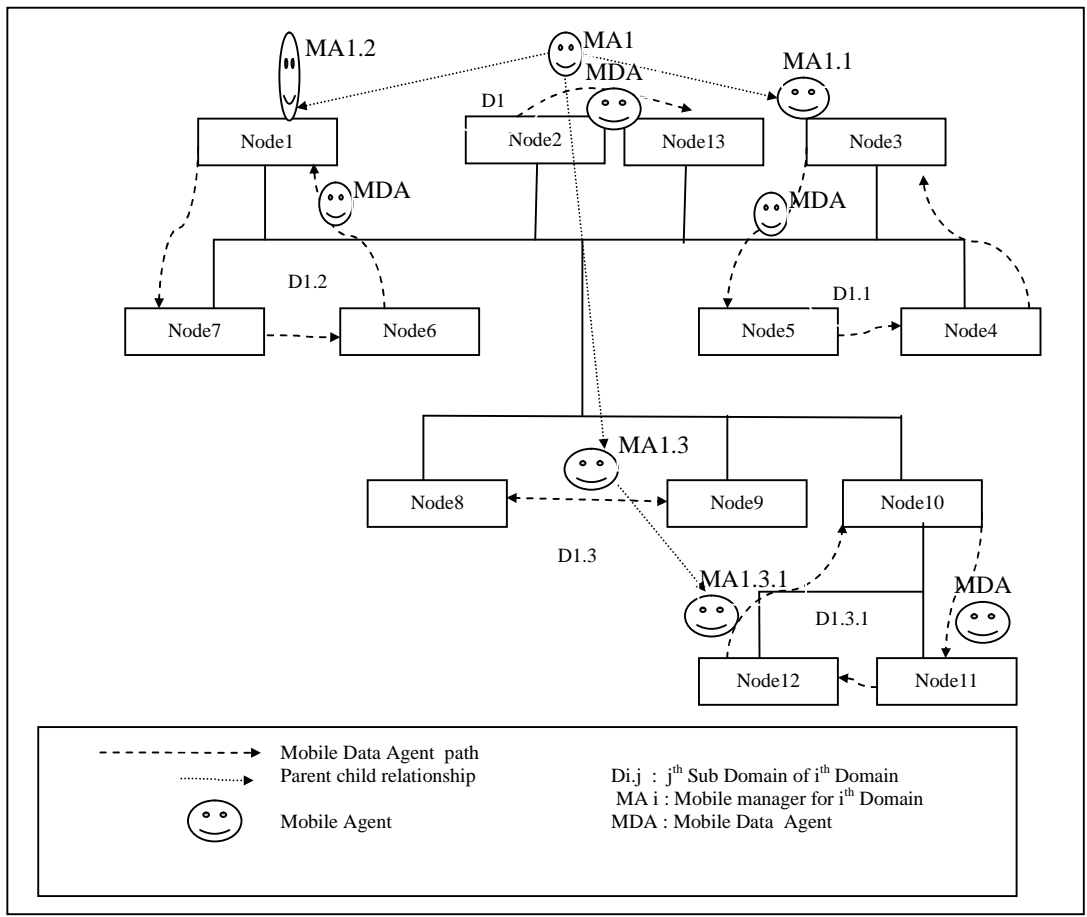

Figure 7. Managed network View after addition of Node 11-12 into domain D1.3

Once created and deployed in their sub-domains, M-SNLMs manage all of the network management activities of their domains. For all practical purposes, they manage the complete management traffic of their assigned domains. All configuration management tasks like downloading of default configuration data, managing discovery of new devices, fault reports, performance monitoring reports etc. are managed by these managers only. They communicate only the aggregate reports or data to their supervisors. This could be either polling based or asynchronously through events.

The mother Manager can interact with its child manager at any time and view the status of all of its child managers. The communication between mother and child manager is based on communication events. Thus manager at higher level can view status of all the domains nodes which are under the controls of its lower level managers.

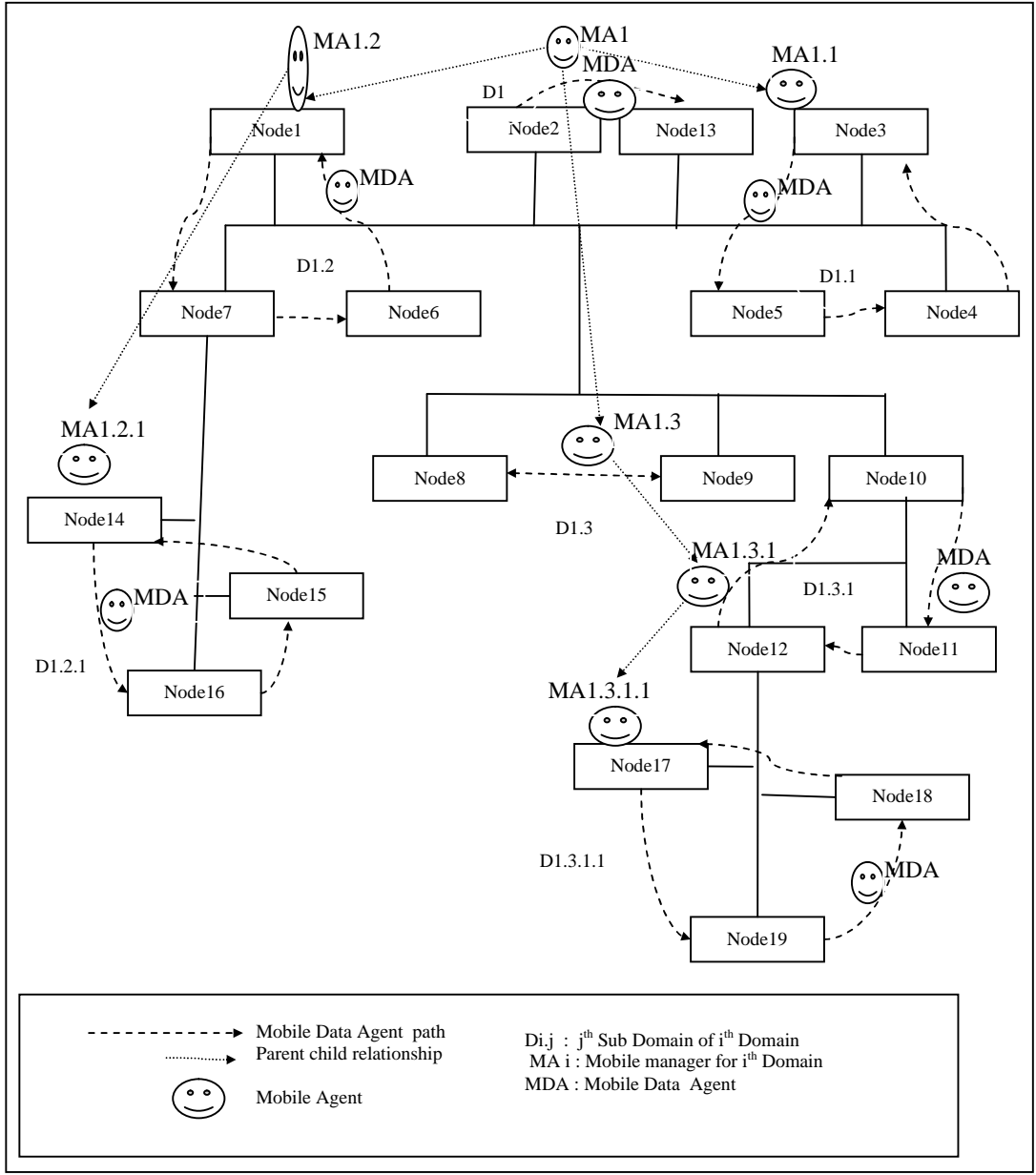

Figure 8.  Managed network View after addition of Node 13, Node14-16 and Node 17-19 into domains D1, D1.3.1 and D1.2 respectively

## 4. NETWORK MANAGEMENT COST CALCULATION

Network management traffic i.e. get_request & set_request passing through system network have some cost associated with it. This cost can be in terms of bandwidth required for management data transfer or in term of the time for which network is not free for user's data use[18][19][20]. Each model has different cost for network management traffic. It is one of research topic that how to minimize the network management cost. The cost for network management does not only depends on management data size but also depends on cost coefficient of network link through which management data pass. To calculate the cost of the IMASNM model, following cost calculation considerations were taken into account.

### 4.1. Network Management Cost for Centralized Client/Server (C/S) Model Based on SNMP

For Client/Server based network management model cost of polling N network devices is

$$C_{c/s} = \sum_{i=1}^{N} K_{0,i} * (S_{req} + S_{res}) \qquad \ldots\ldots\ldots 5.1$$

Where $K_{0,i}$: Cost coefficient of manager ($0^{th}$ location) to managed node$_i$ link,
$S_{req}$: Size of SNMP request packet, $S_{res}$: Size of SNMP response packet

It may be noted that the term $S_{req} + S_{res}$ represents the data flowing through the link. If p is number of times polling done for a time interval, say one hour, then cost for network management for that time interval is

$$C_{c/s} = (\sum_{i=1}^{N} K_{0,i} * (S_{req} + S_{res})) * p \qquad \ldots\ldots\ldots 5.2$$

### 4.2. Network Management Cost for Mobile Flat Bed Model

For Mobile Agent based flat bed Network management model [19][20] cost of one round trip through a network of N+1 nodes (with N0 acting as central manager node) is

$$C_{MAFB} = K_{0,1} * S_{MA} + K_{1,2} * (S_{MA} + D) + K_{2,3} * (S_{MA} + 2D) + K_{3,4} * (S_{MA} + 3D) + \ldots + K_{N-1,N} * (S_{MA} + (N-1)D) + K_{N,0} * (S_{MA} + ND)$$

$$C_{MAFB} = \{ \sum_{i=0}^{N-1} K_{i,i+1} * (S_{MA} + i*D) \} + K_{N,0} * (S_{MA} + ND) \qquad \ldots\ldots\ldots 5.3$$

Where $K_{i,j}$: Cost coefficient of link between $i^{th}$ and $j^{th}$ nodes,
$S_{MA}$: Size of mobile agent,
D: Size of information collected by MA from each node. (D is usually the aggregate information of various MIB variables and is quite small in size (bytes) in comparison to raw data collected (MIB variables) in Client/Server mode. This is so as MA agent performs like as a manager locally.) If p is no of times the polling done for a time interval for network management then management cost for that interval is

$$C_{MAFB} = (\{\textstyle\sum_{i=0}^{N-1} K_{i,i+1} * (S_{MA}+i*D)\} + K_{N,0} * (S_{MA}+ND))*p \ \dots\dots\dots 5.4$$

Thus from the two equations 5.1 and 5.2 we have observed that management cost in Client/Server Network management model is directly proportional to number of request & response made to managed a device or the number of times a particular device MIB is accesses to retrieve the information to be manages, whereas as per equations 5.3 and 5.4 in MA based model it is directionally proportional to MA agent size as well as amount of information collected from mobile agent.

## 4.3. Network Management Cost for IMASNM Model

The Network management cost for IMASNM Model is shown in equation 5.5. The IMASNM Model Network management cost involves not only for management traffic cost i.e (messages between the managers and sub-domain managers) but also the cost of setting up the managers as per the initial discovery of the network.

$$C_{IMASNM} = C_{IMASNMD} + C_{IMASNMP} \qquad\qquad \dots\dots\dots 5.5$$

Where $C_{IMASNMD}$: cost for discovering the network and deploying the managers as per IMASNM model.

$C_{IMASNMP}$: cost of a typical polling to know whole network status at top most level.

The cost of setting up the top level managers as per the initial discovery of the network would be

$$C_{IMASNMD} = \textstyle\sum_{h=1}^{L} \ \textstyle\sum_{j=0}^{M-1} F_{h,j} * MA_{Size} \qquad\qquad \dots\dots\dots 5.6$$

Where $MA_{Size}$: Size of mobile manager,

L: Number of mother manager in the network,

M: Number of child managers in the sub domains of $h^{th}$ mother manager and

$F_{h,j}$: Sum of all the link cost coefficient between $h^{th}$ mother manager to $j^{th}$ child manager's sub domain.

$$C_{IMASNMP} = \textstyle\sum_{h=1}^{L} \ \textstyle\sum_{j=0}^{M-1} F_{h,j}(MA_{res}) + \textstyle\sum_{j=1}^{Q} C_{Q} \qquad\qquad \dots\dots\dots 5.7$$

Where $MA_{res}$: Size of message sent by child to mother manager to report network health of its underlying domain,

$C_{Q}$: Flat bed model cost for $Q^{th}$ domain's M-SNLM.

$$C_{Q} = MDA_{Size} * (R_{Q}+1) * K_{Q} \qquad\qquad \dots\dots\dots 5.8$$

Where $MDA_{Size}$: Size of mobile data agent, $R_{Q}$ is number of managed node is $Q^{th}$ domain and $K_Q$ is link coefficient of link's of $Q^{th}$ domain. If p is no of polling done for a time interval for network management then management cost for that interval is

$$C_{IMASNM} = (C_{IMASNMD} + C_{IMASNMP}) *p \qquad \ldots\ldots\ldots 5.9$$

Where $C_{IMASNMD}$: One times deployment cost so it can be ignored for long term analysis of network health.

$C_{IMASNMP}$: Cost for IMASNM for each polling interval.

## 5. QUANTITATIVE EVALUATION AND COMPARISON OF NETWORK MANAGEMENT COST FOR IMASNM MODEL AND CLIENT SERVER NETWORK MANAGEMENT MODEL

In order to calculate the respective cost, we consider a network station that has a number of nodes in its management domain as shown in figure 9. To perform some management task, the management station needs to check the MIB in every node in its domain and other sub-domains, one after another, according to certain algorithm. The network management cost for both the models is calculated in terms of amount of data transfer through network links for management activities i.e to know the health of network.

The framework for mobile agent based network management is described in paper [21]. The size of data transfer is calculated using "SoftPerfect Network Protocol Analyzer" sniffer tool [22]. Aglet Software development kit is (ASDK2.0.2) [23] used as Mobile agent platform [17] for implementing NM framework. SNMP Tool Kit, WebNMS Agent Toolkit Java [24] Edition 6 (Advent Net 6.0) [25] is used to manage low level i.e. device's information. This information is required for network management application to analyse network status.

For Client/Server Model size of $SNMP_{req}$ and $SNMP_{res}$ message for a single MIB variable of managed object is 83 byte and 84 byte respectively. The cost coefficient $K_{i,j}$ is taken as 1, if both i & j belong to same domain and 5 if they are across the domains.

The cost for the Client/Server Model, as per equation 5.1, for the network shown in figure 9 to retrieve five parameters (X1-X5) from MIB at each node would be as shown below. Node 3 in domain D1 is taken as the network management station node or central node.

$C_{C/S} = \sum_{i=1}^{18} K_{3,i} *(83+84) *5$

$=835*(K_{3,1}+K_{3,2}+K_{3,3}+K_{3,4}+K_{3,5}+K_{3,6}+K_{3,7}+K_{3,8}+K_{3,9}+K_{3,10}+K_{3,11}+K_{3,12}+K_{3,13}+K_{3,14}+K_{3,15}+K_{3,16}+K_{3,17}+K_{3,18})$

$=835(1+1+0+5+6+6+6+6+5+10+11+11+11+11+10+10+11+11)$

$=835*132$

$=110220$ byte

If p=20 polling interval in an hour the management cost for an hour is (as per equation 5.2)

$C_{C/S} = 110220*20$

$=2204400$ byte

The cost for IMASNM model, as per equations 5.5 to 5.7, given the size of MDA=3.2kb, size of MA =3.92kb and size of $MA_{res}$= 583byte, then network management cost for the network shown in figure 9.

$C_{IMASNMD} = 3.92*1024*(F_{1,1.2}+F_{1,1.1}+F_{1.2,1.2.1}+F_{1.2,1.2.2}+F_{1.1,1.1.1})$

$=4014.08*(5+5+5+5+5)$

$=100352$ byte

It is a onetime deployment cost for M-SNLMs and can be ignored for long term polling jobs.

$C_{IMASNMP} = (583)*(F_{1,1.2}+F_{1,1.1}+F_{1.2,1.2.1}+F_{1.2,1.2.2}+F_{1.1,1.1.1}) + 6*3.2*1024*3$

$=583*25+6*3276.8*3$

$=14575+58980$

$=73555$ byte

$C_{IMASNM}$    73555 byte (Not accounting the deployment cost)

If p = 20 polling interval in an hour the management cost for an hour is

$C_{IMASNM} = C_{IMASNMP} * 20$

$= 73555 * 20$

$=1471100$ byte

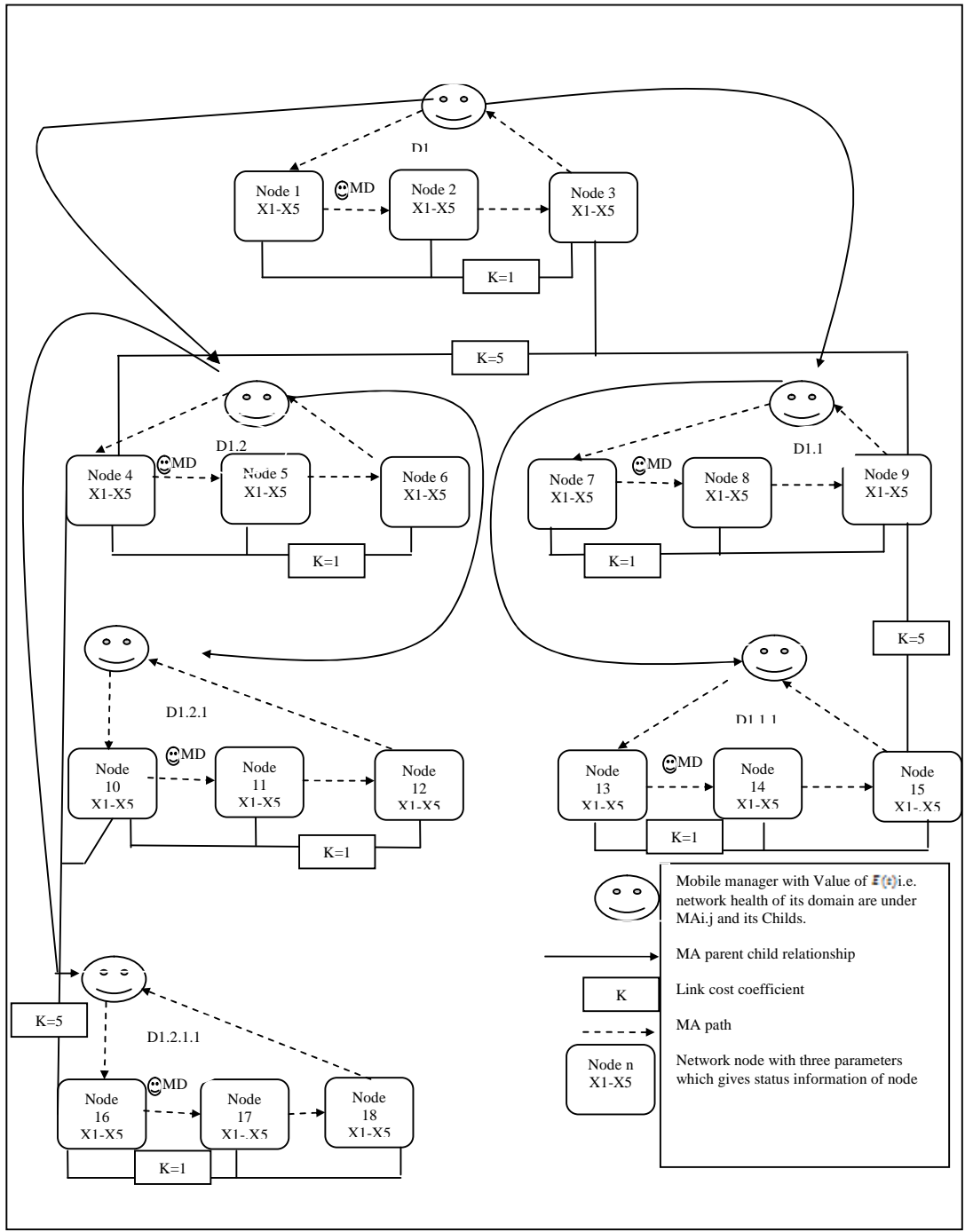

Figure 9. Managed network View For network management cost calculation

The Network management cost comparison between Client/Server and IMASNM with respect to polling intervals for Network health is given in table1 and graph in figure 10.

**Table 1. Network Management cost Comparison between C/S and IMASNM models**

| Number of Polling | $C_{C/S}$ (Kb) | $C_{IMASNM}$ (Kb) |
|---|---|---|
| 1 | 110.22 | 73.55 |
| 10 | 1102.21 | 735.55 |
| 20 | 2204.41 | 1471.11 |
| 50 | 5381.84 | 1769.63 |
| 100 | 10763.67 | 3441.26 |

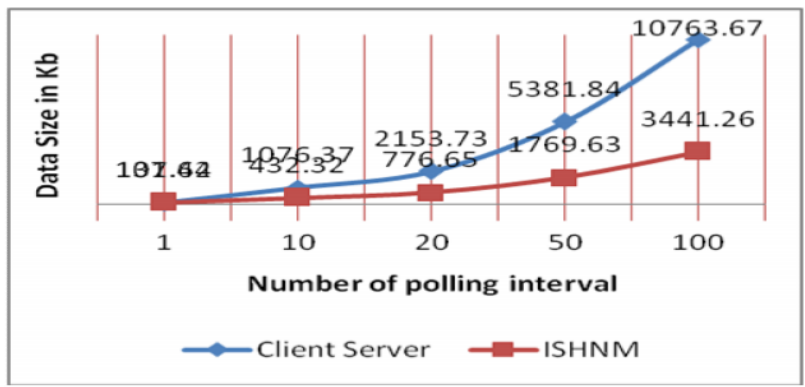

Fig.10. Network Management cost Comparison between C/S and IMASNM models

In client server model the management cost in terms of the data transferred to the Global manager for the whole network is directly proportional to the following factors:

1. Number of requests and responses to fetch the data remotely

2. Cost coefficient of the links on which information is exchanged.

Whereas in the proposed IMASNM model, M-SNLMs manage the domains locally thereby minimizing the cost incurred due to costly inter-domain link traversal. Additionally the cost of managing flat bed model, being managed under each M-SNLM, is less than the cost of managing the sub-network of the same size via client server model.

Therefore it may be noted from Table 1 and Figure 10 that management cost in terms of data transferred to the Global Manager (central manager) for the whole network is drastically less in case of proposed IMASNM model as compared to the client server model where it grows exponentially.

## 6. CONCLUSION

As the next generation data and telecommunication networks become more complex and distributed in nature, typical centralized network management systems (SNMP and CMIP) based on client server paradigm, suffer from scalability and flexibility problems as it involves the transmission of large amount of management data towards the centralized management station for processing. A distributed paradigm is a viable solution to perform management functions when network start growing significantly. In this sense, a dynamic intelligent scalable management model based on mobile agents provides an attractive perspective to the lack of flexibility and scalability of current centralized management systems.

This paper presents a feasible model and architecture (IMASNM), deploying mobile agents, for efficient network management. The proposed model discusses strategies for large scale network partitioning, fixing of management scope and deployment of mobile M-SNLMs in various sub-network domains. With a view to manage scalability the proposed model not only defines strategy to spawn child M-SNLM and their placement in sub-network domains but also the communication mechanism with higher level managers. Furthermore the management scalability has been improved by placing M-SNLMs close to the sub-network to be managed and only returning needed management data to the supervisors.

The experimental results establish, from the cost analysis, that the proposed mobile agent based architecture (IMASNM) scales better than client server model.

# REFERENCES

[1]. Mani Subramanian,"Network Management: Principles and Practice", 2008 Edition, Pearson Education

[2]. Makki, S., Wunnava, S., 2008. Next Generation Networks and Code Mobility. ISAST Transactions on Communication and Networking, No.1, Vol 2

[3]. "Agent based computing: A booklet for executing", http://www.eurescom.de/

[4]. Stallings, W., 1999. SNMP, SNMPv2, SNMPv3 and RMON 1 and 2, third ed. Addison Wesley

[5]. Yemini, Y., 1993. "The OSI Network Management Model", IEEE Communications Magazine, vol.31, no.5, pp.20-29. May.

[6]. Stephan, R., Ray, P., Paramesh, N., 2004. Network management platform based on mobile agents. International Journal of Network Management 14, 59–73.

[7]. Liotta, A., Pavlou, G. Knight, G., 2002. "Exploiting agent mobility for large-scale network monitoring", IEEE Networks, 16:7-15.

[8]. Rubinstein, M.G., Duarte, O., Pujolie, G., 2002. "Scalability of a network management application based on mobile agents", Journal of Communication and Networks, 5:240-248.

[9]. Manvi, S.S., Venkataram, P., 2006. Agent based subsystem for multimedia communications. IEEE Proceedings Software Journal 153 (1), 38–48.

[10]. Mydhili K Nair, Gopalakrishna V, 2011, Applying Web Services with Mobile Agents For Computer Network Management, International Journal of Computer Networks and Communications (IJCNC), Vol. 3, No. 2, pp 125-144, March.

[11]. Gavalas, D.,Tsekouras, G.E., Anagnostopoulos, C., 2009, A mobile agent platform for distributed network and system management, Journal of Systems and Software, Vol 82, pp. 355-371

[12]. Satoh, I., 2006. Building and selecting mobile agents for network management. International Journal of Network and Systems Management 14 (1), 147–169. March.

[13]. Goldszmidt G., Yemini Y., Yemini S., 1991. "Network management by delegation", Proceedings of the 2nd Int. Symposium on Integrated Network Management (ISINM'91), April.

[14]. SNMP research, "The Mid-Level Manager", http://www.snmp.com/products/mlm.html

[15]. Baldi, M., Gai, S., Picco, G.P. 1997. "Exploiting Code Mobility in Decentralized and Flexible Network Management", in Proceedings of First International Workshop on Mobile Agents (MA'97), pp. 13-26, Berlin, Germany.

[16]. Khan, B. Kleiner, D.D. Talmage, D., 2001. OPTIPRISM: a distributed hierarchical network management system for all-optical networks, Global Telecommunications Conference, GLOBECOM '01. IEEE.

[17]. Leila, Ismail, 2008, A Secure Mobile Agent Platform, Journal of Communication, vol 3, No 2, pp 1-12

[18]. Jian Ye, Symeon Papavassiliou, 2008. An analytical framework for the modelling and evaluation of the mobile agent based distributed network management paradigm. International Journal of High Performance Computing and Networking, vol. 5, pp. 273-284

[19]. Al-kasassbeh, M., Adda, M., 2008. Analysis of mobile agents in network fault management Journal of Network and Computer Applications, 31 (4), pp. 699–711

[20]. Gavalas, D., Greenwood, D., Ghanbari M. and O'Mahony, M. 2002. "Hierarchical network management: A scalable and dynamic mobile agent-based approach." Computer Networks, 38: 693-711, 2002.

[21]. Singh, V., Mishra, A., and Sharma, A.K. 2010. "A Mobile Agent Based Framework for Network Management System", International Conference on Computer Engineering and Technology, ICCET' 2010

[22]. SoftPerfect: A Network Protocol Analyser, www.**softperfect**.com/products/**networks**niffer/, 2009

[23]. IBM Corp, The Aglets Home http://www.trl.ibm.co.jp/aglets/, 2009.

[24]. JAVA, http://www.java.com, 2009

[25]. WebAdventNet SNMP API 4, http://www.webnms.com, 2011.

**Author**

**Prof. A. K. Sharma**[1] received his M.Tech. (Computer Sci. & Tech) with Hons. from University of Roorkee (now IIT, Roorkee) in the year 1989 and Ph.D (Fuzzy Expert Systems) from JMI, New Delhi in the year 2000. From July 1992 to April 2002, he served as Assistant Professor and became Professor in Computer Engg. at YMCA Institute of Engineering Faridabad in April 2002. He obtained his second Ph.D. in IT from IIIT & M, Gwalior in the year 2004. His research interests include Fuzzy Systems, Object Oriented Programming, Knowledge Representation and Internet Technologies. He has successfully guided 12 Ph.D. thesis.

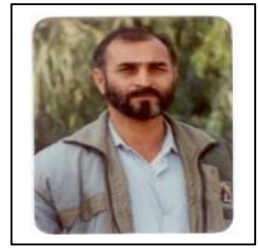

**Atul Mishra**[2] is working as an Associate Professor in the Department. of Computer Engineering, YMCA University of Science and Technology, Haryana, India. He holds a Masters Degree in Computer Science and Technology, from University of Roorkee (now IIT, Roorkee) and is currently pursuing his PhD in Computer Science and Engg. from MD University, Rohtak. He has about 16 years of work experience in the Optical Telecommunication Industry specializing in Optical Network Planning and Network Management tools development. His research interests include SOA, Network Management & Mobile Agents.

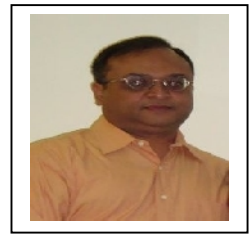

**Vijay Singh**[3] received his M.Tech. (Computer Engineering) from from MD University, Rohtak in the year 2010. His research interests include Object Oriented Programming, Network Management and Mobile agents.

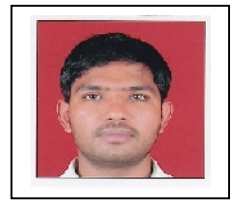